\documentstyle[sprocl]{article}

\input{psfig}

\def\be{\begin{equation}}

\def\ee{\end{equation}}

\def\bea{\begin{eqnarray}}

\def\eea{\end{eqnarray}}

\begin{document}

\begin{flushright}
NUB-TH-3147/96\\
CTP-TAMU-52/96\\
\end{flushright}

\title{SUPERSYMMETRIC DARK MATTER}
\author{ PRAN NATH }
\address{Department of Physics, Northeastern University, 
Boston\\ MA 02115, USA}
\author{ R. ARNOWITT}
\address{Center for Theoretical Physics,Department of Physics,Texas A $\&$ M
University, College Station\\ TX77843, USA}

\maketitle\abstracts{ A review of recent developments in the analyses of
supersymmetric dark matter is given. }

\section{Introduction}

Currently there is  a great deal of evidence  for the existence of
dark matter in the universe: in our galaxy, in other galaxies and
in galactic clusters, as  much as 90$\%$ of matter in the universe
may be dark\cite{kolb,smith,jungman,primack}. There are many condidates for dark matter in astronomy such
as jupiters, brown dwarfs, blackholes, etc. However, not all of dark matter 
can be baryonic and thus  a number of particle physics candidates have been
suggested.These include  massive neutrinos, axions, supersymmetry candidates
as well as exotic candidates such as topological defects. A convenient 
parametrizaton of matter in the universe is  given by the quantity,
$\Omega$=$\rho/\rho_c$, where $\rho$ is the matter density of the universe
and  $\rho_c$= 3 $H^2/8\pi G_N$ is the critical matter needed to close 
the universe, and H is the Hubble parameter which is generally parametrized 
by H=100 h km/s Mpc. Experimentally the current evaluations of h lie in the 
interval 0.4 ${\stackrel{<}{\sim}}$ h ${\stackrel{<}{\sim}}$0.8.
An attractive hypothesis theoretically is that of an inflationary universe
with $\Omega=1$. The quantity that enters in theoretical analyses of dark
	matter is $\Omega h^2$. The COBE data indicates that there are more
	than one component to the non-baryonic dark matter, e.g., a hot component
	comprised of particles 
	which would be relativistic at the time of galaxy formation
	 and a cold component comprised of particles 
	which would be non-relativistic  at the time of galaxy formation.
	The hot component could be a massive neutrino
	 and the cold component could be either axions,
	or a susy candidate. Actually, in addition to the cold-hot 
	dark matter fit to the power spectrum data observed by COBE
	and other experiments 
	there are other possibilities as well, such as inclusion of 
		 a cosmological constant. We list  below some examples:
		(i) CHDM (cold-hot DM with $\Omega_{\tilde\chi_1^0}$ = 0.7,
$\Omega_{HDM}$ = 0.25, $\Omega_B$ = 0.05 (baryonic DM), h = 0.5) giving
$\Omega_{\tilde\chi_1^0}~h^2$ = 0.18;
 (ii) $\Lambda$CDM
(cosmological constant DM with $\Omega_{\tilde\chi_1^0}$ = 0.25--0.35,
$\Omega_B$ = 0.05, $\Omega_{\Lambda}$ = 0.6--0.7, h =0.8) with
$\Omega_{\tilde\chi_1^0}~h^2$ = 0.16--0.23.;
 (iii) tCMD (tilted DM with
$\Omega_{\tilde\chi_1^0}$ = 0.95, $\Omega_B$ = 0.05, h = 0.5--0.6, n = 0.8)
giving $\Omega_{\tilde\chi_1^0}~h^2$ = 0.24--0.34. In the analysis of this
review we shall assume that the  parameter space is limited by 
~\\
\begin{equation}
0.1\leq\Omega_{\tilde\chi_1^0}~h^2\leq 0.4
\end{equation}
~\\ 
 We will discuss later on the effect of varying the allowed range of
 $\Omega_{\tilde\chi_1^0}~h^2$.

Theoretically the analysis of $\Omega_{\tilde\chi_1^0}~h^2$ is interesting as
it involves several different fields of physics: 
particle physics enters via the fundamental interactions of the dark
matter candidates, cosmology enters as the
relic density involves both the gravitational constant $G_N$ and the 
 Hubble parameter H, and statistical physics enters via the
 appearance of the Botzmann  constant k in the analyses of relic density
 in the early universe. Further, nuclear physics enters as one considers
 specific detection techniques such as scattering of dark matter particles
 from nuclei.
 
	In this review our focus will be to explore in detail 
	the supersymmetric candidates for cold dark matter.  
	In  the Minimal Supersymmetric Standard Model (MSSM)
	 there are two  candidates  which
		are  neutral and have odd R parity: 
		these are the neutralino and the 	
	sneutrino. If we assume that one of these is the lowest
	mass supersymmetric particle (LSP) and further assume 
		that R parity is conserved then, one of the
		candidates could be a CDM candidate.  
	Unfortunately there are 110 arbitrary parameters in MSSM,
	and thus the masses of the supersymmetric particles are all
		arbitrary. Thus computations of 
		supersymmetric dark matter in the framework of 
		MSSM are by their very nature highly unpredictive.
	In contrast, in the minimal supergravity grand 
	unification\cite{cham,hlw}
	 one can describe
	low energy physics in terms of just four parameters. Thus the
	theory is very predictive. 
		
		A topic of great interest is the laboratory detection 
		of dark matter. In this context the scattering of dark
		matter particles from nuclei appears very promising and
			we will devote considerable amount of discussion
			to this topic in this paper.
			The outline of this paper is as follows: In Sec 2 
			we discuss  dark matter in supergravity unification;
			in Sec 3 a discussion of the relic density analysis
			is given; 
			in Sec 4  we discuss  constraints on dark matter 
			analyses;
			Sec 5 is devoted to a discussion of the
			detection of dark matter; in Sec 6 we give an
			analysis of event rates for the case of universal 
			soft SUSY breaking parameters; in Sec 7 we discuss
			the effects of  non-universal breaking; in Sec 8
			an analysis of event rates with non-universal 
			soft SUSY breaking is given and in Sec 9 we give
			conclusions. 
			
\section{~~Dark Matter in Supergravity Grand Unification}
\indent 
As discussed above the MSSM is very 
unpredictive. The reason for this is that there is no 
phenomenologically acceptable mechanism for the breaking of supersymmetry 
in the MSSM. Consequently one must add by hand the soft SUSY breaking 
parameters and there are as many as 110 arbitrary parameters that
one can add in the soft SUSY breaking sector. For this reason we shall
work here within the framework of supergravity grand unification which has 
a much more limited parameter space. The models we have in mind are those 
where suspersymmetry is broken in the hidden sector and communicated to
the physical sector by gravitational interactions, and we consider here 
the minimal model. We assume that the 
basic theory in the physical sector  is a grand unified theory with a 
GUT group G which breaks at a scale $M_{GUT}\simeq 10^{16}$  to 
 SU(3) x SU(2) x U(1). After the 
breaking of supersymmetry and the GUT symmetry, and after integration over
the fields of the hidden sector and the superheavy fields, one finds the
effective potential  in the SUSY breaking sector to be of the form:
\begin{equation}
V_{SB}=m_{o}^{2}\Sigma_{a}z_{a}^{+}z_{a}+(A_{o}W_{Y}+B_{o}\mu_{o}H_{1}H_{2}+h.c)
\end{equation}
\begin{equation}
L^{\lambda}_{mass} =-m_{1/2}\bar{\lambda}^{\alpha}\lambda^{\alpha}
\end{equation}
\noindent
where $\{z_{a}\}$ are the scalar fields , $\lambda^{\alpha}$ the gaugino
fields, $\mu$ is the Higgs mixing parameter and $W_Y$ is the cubic 
Yukawa coupling which enters in the  effective superpotential,W, of the low
energy theory as follows:
\begin{equation} 
W=\mu_{o} H_{1}H_{2}+W_{Y} + {1\over{M_{G}}}W^{(4)} 
\end{equation}
\noindent
 Here $W^{(4)}$ contains any quartic
non-renormalizable couplings which may lead to proton decay. At this level
the theory contains the four SUSY breaking parameters
$m_0$, $m_{1/2}$, $A_0$, and $B_0$ in addition to the Higgs mixing
parameter $\mu_0$, and the GUT parameters $M_G$ and $\alpha_G$.

	The dynamics of supergravity grand unification  breaks the 
	electro-weak symmetry dynamically by radiative 
	effects\cite{inoue,ellis1,ross}. Further,
	one typically uses the radiative breaking of the electro-weak 
	symmetry to  compute the Higgs mixing parameter $\mu$ by fixing
	the Z mass, and one eliminates the parameter B(which is the value
	of the parameter $B_0$  at the electro-weak scale) in favor of\\
	tan$\beta\equiv <H_2>/<H_1>$ such that

\begin{equation}
\mu^{2}={{\mu_{1}^{2}-\mu_{2}^{2}tan^{2}\beta}\over{tan^{2}\beta -1}}-{1\over
2}M_{Z}^{2};~~ sin2\beta = {{-2B\mu}\over{2\mu^{2}+\mu_{1}^{2}+\mu_{2}^{2}}}\\
\end{equation}
\noindent
Here  $\mu_{i}^{2}=m_{H_{i}}^{2}+\Sigma_{i}$ where $m_{H_{i}}^{2}$ is the
running $H_{i}$ mass at scale Q $\approx M_{Z}$ and $\Sigma_{i}$ are loop
corrections.  The $m_{H_{i}}^{2}$ are given by
\begin{equation}
m_{H_{1}}^{2}=m_{o}^{2}+m_{1/2}^{2} g(t)\\
\end{equation}
\begin{equation}
m_{H_{2}}^{2}=m_{1/2}^{2}e(t)+A_{o}m_{1/2}f(t)+m_{o}^{2}h(t)-k(t)A_{o}^{2}\\
\end{equation}
\noindent
where the form factors e,f,g,h,k are defined in ref(10)
 and t=ln($M_{G}^{2}/Q^{2}$).  One may
show that solutions exist to Eqs. (5), i.e. that SU(2)$\times$U(1) is
spontaneously broken, if and only if at least one of the supersymmetry soft
breaking interactions are non-zero.  Thus it is the supergravity 
interactions at $M_{G}$ that give rise to the breaking of
SU(2)$\times$U(1) at the electroweak scale.
 The $SU(3)\times SU(2)\times U(1)$ gaugino masses at the electro-weak scale are given 
by

\begin{equation}
m_{i}=(\alpha_i/\alpha_{G})m_{1/2} 
\end{equation}

  In the above supergravity grand unification the sparticle mass spectrum is 
  	determined dynamically in terms of the four parameters, and thus
  	the theory predicts what the LSP is. It is quite remarkable that
  	over most of the parameter space one finds that the LSP is indeed
  	a neutralino  ${\tilde\chi_1^0}$.  Thus  supergravity predicts 
  	the ${\tilde\chi_1^0}$ to be the cold dark matter of the universe.
The $\tilde{\chi}_{1}$ is of course an admixture of gauginos $\tilde{W}_{3},
\tilde{B}$ and Higgsinos $\tilde{H}_{1}, \tilde{H}_{2}$ and one may write
\begin{equation}
\tilde{\chi}_{1}=n_{1}\tilde{W}_{3}+n_{2}\tilde{B}+n_{3}\tilde{H}_{1}+n_{4}
\tilde{H}_{2}\\
\end{equation}
\noindent
where we  use  the convention that $\tilde{\chi}_{1}$ is the lightest 
neutralino and   
where $n_{i}$ are the components of the gaugino states and are to be 
determined  by diagonalizing the
neutralino mass matrix. 
The neutralino mass matrix in the basis 
($\tilde{W}_{3}, \tilde{B}, \tilde{H}_{1},
\tilde{H}_{2}$) is given as follows:

\begin{equation}
M_{\tilde{Z}}=\pmatrix{\offinterlineskip
{\tilde{m}_{2}}&o&\vrule\strut&a&b\cr
o&{\tilde{m}_{1}}&\vrule\strut&c&d\cr
\noalign{\hrule}
a&c&\vrule\strut&o&{-\mu}\cr
b&d&\vrule\strut&{-\mu}&o\cr}
\end{equation}

\noindent
Here  $a=M_{Z}cos\theta_{W} cos\beta$,~ $b=-M_{Z} cos\theta_{W} sin\beta$,~
$c= -M_{Z} sin\theta_{W} cos\beta$ and $d=M_{Z} sin\theta_{W} sin\beta$. 
In general one needs a numerical analysis to determine the value of the
mass eigenvalues as well as the eigenvectors. In the limit 
 $\mu^{2}>>M_{Z}^{2}$, which holds in a significant part of the parameter
 space, it is posssible to obtain an analytic form for the indices 
 $n_i$ as an expansion in  $\mu/M_Z$. The results are given in 
 appendix A. This expansion is useful in understanding the phenomena
 associated with the neutralino relic densities and event rates.
 Generally one finds that in this limit the neutralino is mostly a
 Bino with  $n_{2} > 0.95$ and often larger, while 
$n_{3}, n_{4}$ and $n_{1}$ are  typically of first order, $O(M_{Z}/\mu)$. 
 Because of this the Higgsino components are not totally negligible, 
 and they are often quite significant e.g.
$n_{3}\approx 0.2$.  Hence one cannot neglect the 
Higgsino components even when the neutralino  may be dominantly a 
Bino. This result has important implications in  the analysis of
event rates. \\

Before proceeding further we wish to discuss the effect
that  a heavy top quark has on the electro-weak breaking.
One of the   major effects of a heavy top arises because
it lies close to the Landau pole value which is given by 
  $m_t^f\approx 197 sin\beta$. This occurs  because  the
proximity of the top to the Landau pole  drives the lighter stop towards
a tachyonic limit.  We exhibit this phenomenon in some detail. 
The mass of the lighter stop $\tilde t_1$ is determined by the diagonalization
of the following stop $(mass)^2$ matrix:

\begin{equation}
\left(
{{ {m_{\tilde t_R}^2}\atop{m_{t} (A_t - \mu ctn \beta)}}
{{m_t (A_t - \mu ctn \beta)} \atop {m_{\tilde t_L}^2}   }}
\right)
\end{equation}
The lighter stop (mass$)^2$ 
eigen-value in the vicinity of the Landau pole is given by\cite{nwa}

\begin{equation}
m_{\tilde{t}_{1}}^2=-{1\over 3}{A_R^2\over D_o}+m_{\tilde{t}_{1}}^2(NP)
\end{equation}
\noindent
where
\begin{equation}
A_R\cong A_t-0.613 m{\tilde{g}};~~ D_o\cong 0.164
\lbrack(m_t^f)^2-(m_t)\rbrack^2/M_W^2 
\end{equation}
\noindent
 and where $m_{\tilde{t}_{1}}^2$
(NP) is a  non-pole contribution.
  The condition that  ${\tilde{t}_{1}}$  not turn tachyonic limits 
severely the allowed range of $A_t$. 

\section{Relic Density Analysis}
An important quantity in the analysis of supersymmetric dark matter is the 
LSP relic density. After the time of the Big Bang, the primordial neutralinos
begin to annihilate into ordinary matter, e.g., via the process
${\tilde\chi_1^0}+{\tilde\chi_1^0}\rightarrow f+
\bar{f},\ldots$, where $f, {\bar f}$ are leptons, quarks, etc\cite{lee}.
 The annihilation  is governed by the Boltzman
equation :
\begin{equation}
\frac{dn}{dt}=-3Hn-<\sigma v>(n^2-n_0^2)
\end{equation}
where n = number/vol is the $\tilde\chi_1^0$  number density,
$n_0$ is the value of n at thermal equilibrium, 
 H is the Hubble parameter at time t, $\sigma$ is the neutralino 
 annihilation cross section, v is the relative neutralino velocity,
  and $<\sigma v>$ stands for the thermal average.   
 Assuming that the neutralinos  are non-relativistic at the
 time of interest (a realistic approximation)
 one has 
\begin{equation}
<\sigma v>=\int_0^{\infty} dvv^2\sigma v~exp[-v^2/4x]/\int_0^{\infty} 
dvv^2 exp[-v^2/4x]
\end{equation}
\noindent
 
At large temperatures close  to the Big Bang the  $\tilde\chi_1^0$ is
in thermal equilibrium with the background. However, the
$\tilde\chi_1^0$ decouples from the background when the annihilation 
rate becomes smaller than the expansion rate given by the Hubble term. 
This decoupling takes place at the "freezeout" temperature $T_f$. 
Integration of the Boltzman equation from the freezeout to the 
current temperatures gives the current relic density. One finds 
 ~\\
\begin{equation}
\Omega_{\tilde\chi_1^0} h^2\cong 2.48\times 10^{-11}{\biggl (
{{T_{\tilde\chi_1^0}}\over {T_{\gamma}}}\biggr )^3} {\biggl ( {T_{\gamma}\over
2.73} \biggr)^3} {N_f^{1/2}\over J ( x_f )}
\end{equation}
~\\
\noindent
where
~\\
\begin{equation}
J~ (x_f) = \int^{x_f}_0 dx ~ \langle~ \sigma \upsilon~ \rangle ~ (x) GeV^{-2}
\end{equation}
~\\
\noindent
and  $x_f= kT_f/m_{\tilde{\chi}_{1}}$.  
Now numerical analyses show that freezeout occurs for values of 
$x_f\sim \frac{1}{20}$ which indeed corresponds to the 
neutralinos being  non-relativistic at the time of the freezeout.
The central quantity in the analysis of the relic density is 
$J(x_f)$. The computation of this quantity  is quit subtle as it 
involves a double integral over $(\sigma v)$ which in some regions is 
rapidly varying. An approximation that has generally been made 
is to expand  $\sigma v$ in a power expansion in v,
i.e. $\sigma v\cong a+b(v^2/c^2)+\cdots$. With this approximation the 
computation of 
J($x_f$) becomes trivial. However, it is known\cite{greist}
 that such an expansion
is not a good approximation when ($\sigma v$) is rapidly varying. 
This is  precisely the situation that occurs in the annihilation
of the neutralinos when one is close to the Z and the Higgs 
poles\cite{an1,na0,baer1}.
In this case one must carry out a thermal averaging keeping the
pole structure intact. To circumvent the problem of the double
integration over the pole,  one can carry out analytically one
of the integrations reducing  J($x_f$)  to a single integration
which is computationally easy to evaluate. Details are given in Appendix B.

\section{Constraints on Dark Matter Analyses}

The analysis of dark matter relic density is subject to many constraints. 
Since the analysis is carried out in the framework of supergravity grand 
unification one computes masses of 32 supersymmetric particles in 
terms of the basic SUSY breaking parameters (which are just four for 
the universal soft SUSY breaking case). The analysis of the mass
spectrum is subject to  following constraints:
(i) color and charge conservation
at  the electro-weak scale,(ii) naturalness constraints, i.e., that
$m_0$,$m_{\tilde g}\leq 1$TeV,(iii) LEP, CDF and DO 
 limits on the mass spectrum. Further, 
the decay $b\rightarrow s+\gamma$ represents an important constraint 
on any new physics beyond the Standard Model.
 Recently CLEO  measured the branching ratio for 
$B\rightarrow X_s+\gamma$ to be\cite{alam}:

\begin{equation}
BR(B\rightarrow X_s\gamma))=(2.32\pm 0.5\pm 0.29 \pm 0.32)\times 10^{-4}
\end{equation}

\noindent
 Combining the errors in quadrature gives BR(B$\rightarrow
X_s\gamma)\cong(2.32\pm 0.66)\times 10^{-4}$.In the spectator 
approximation one can relate the above to the flavor changing process
$b\rightarrow s+\gamma$ and one has 
\begin{equation}
{BR(B\rightarrow X_s\gamma)\over BR(B\rightarrow X_ce\bar{\nu}_e)}
\cong{\Gamma(b\rightarrow s+\gamma)\over\Gamma(b\rightarrow
c+e+{\bar{\nu}}_e)}\equiv R
\end{equation}
\noindent
where BR(B$\rightarrow X_c e\bar{\nu}_ e$) = (10.7$\pm$ 0.5).
The  quantity R  turns  out to be  a useful parameter as the factor 
($m_b^5$)(which contains a significant experimental
ambiguity) as well as  some CKM factors  cancel out in the this ratio. 

 The $b\rightarrow s+\gamma$ decay proceeds via loop corrections.
For the case of the Standard Model the loop contribution involves 
a W-t exchange. For the supersymmetric case one has in addition a
$\tilde W -\tilde t$, $H^--t$  and a $\tilde Z -\tilde b$ exchange. 
We see then that
the SUSY contributions enter on an equal footing with the Standard Model
contribution. Thus $b\rightarrow s+\gamma$ is an important constraint on
the SUSY parameter space. 
The supersymmetric contribution to the decay $b\rightarrow s+\gamma$
is governed by effective Hamiltonian\cite{bertolini}

\begin{equation}
H_{eff}=V_{tb}V_{ts}^*{G_F\over\sqrt 2} C_7(M_W)Q_7
\end{equation}

\noindent 
where $F_{\mu\nu}$ is the electromagnetic field strength and 
$Q_7=(e/24\pi^2)$$m_b$$\bar{s}_L$$\sigma^{\mu\nu}$$b_R$$F_{\mu\nu}$.
However,  the computation of  $b\rightarrow s+\gamma$
 must be carried out at the scale $\mu\simeq m_b$ where the decay occurs 
 and thus one must
 use the renormalization group equations to go from the scale
$\mu \approx M_W$ to $\mu\simeq m_b$. Using the above evolution, 
the value of R to leading order (LO) in
QCD corrections is given by
\begin{equation}
R=\mid {V_{tb}V_{ts}^*\over V_{cb}}\mid^2{6\alpha\over \pi I(z)}{\mid
C_7^{eff}(m_b)\mid^2} 
\end{equation}
\noindent
where $I(z)=1 - 8 z^2+ 8z^6-z^8-24 z^4 \ell n z$ is a phase space factor
(z=m$_c/m_b$) and $C^{eff}_7(m_b)$ is given by

\begin{equation}
C^{eff}_7(m_b)= C_2(M_W)+ \eta^{16\over 23} C_7(M_W)+{8\over 3}
(\eta^{14\over 23}-\eta^{16\over 23}) C_8(M_W)
\end{equation}

\noindent
Here  $C_8(M_W)$ is the co-efficient of the operator 
$Q_8=(g_3/16\pi^2) m_b{\bar{s}}_R\sigma^{\mu\nu}T^Ab_L$\\
$G_{\mu\nu}^A$ 
in the renormalization group evolution,
T$^A$ and G$_{\mu\nu}^A$ are the gluon generators and field strengths, and 
  $C_2$
represents the operator mixing with the 4-quark operators in the renormalization
group evolution.

The Standard Model leading order prediction gives\cite{buras} 
$BR[B\rightarrow X_s\gamma]\cong (2.9\pm
0.8)\times 10^{-4}$ for m$_t$ = 174 GeV. Several analyses of 
$b\rightarrow s+\gamma$ in SUSY models within the leading 
order(LO)approximation
have been carried out, and their effects on dark matter 
 analysed. Generally one finds that the 
$b\rightarrow s+\gamma$ constraint is very stringent for $ \mu>0$ 
but less so for the case $\mu<0$. Thus for $\mu>0$ one finds that 
as much as 2/3 of the parameter space is eliminated while for
$\mu<0$ only about 1/5 of the parameter is 
eliminated\cite{na1,borzumati,na2,an3}.
 More recently,
the  next to leading order corrections to the Standard Model have
been computed\cite{misiak}. Analyses including these give results which are 
 similar\cite{baer2} to the ones based on the LO 
 corrections\cite{na1,borzumati,na2,an3}.

\section{Detection of Dark Matter}
 A variety of possibilities for the detection of dark matter which 
 involves more than just its gravitational effects have been discussed
 in the literature.  These include :(i) scattering by 
 nuclei of terrestial detectors ,(ii) annihilation in the centre of 
 sun or earth, (iii) annihilation in the halos of galaxies, (iv) 
 atomic excitations, and (v) superheated microbubbles.  Presently,
 the most promising
 of the above are (i) and (ii), and we shall discuss  specifically (i)
 in detail here. Case(ii) is based on the idea that neutralinos in the
  solar neighborhood  will be attracted to  the center 
 of the sun, where over a period of time they will lose energy and accumulate.
 Annihilation of these will produce neutrinos which in the terrestial  
 detection apparatus will show up as upward moving muons(via their neutral
 current interaction). A similar situation is expected for the
 annihilation in the center of the earth. Background for these processes are
 significantly reduced because of the small angular apertures around the
 center of sun and earth\cite{kamio1,kamio2}. Signals of this type can be 
 picked up by 
 neutrino telescopes.  In case (iii) one expects that 
 neutralinos in the halos  of galaxies will annihilate into 
  $(\gamma, e^+, \bar p, ..)$+X . However, the signal
 to background ratio does not appear very promising in this case.
 Case(iv) involves scattering of neutralinos by electrons bound in atoms
 which may result in atomic excitations\cite{starkman}. 
 Realistic analyses of this 
 possibility are currently lacking. Finally we mention the
 possiblity of superheated microbubbles as a cold dark matter detector
  which has been discussed recently\cite{collar}.

   We turn now to a detailed discussion of neutralino-nucleus 
   scattering\cite{goodman}$^-$\cite{decarlos}.
  The fundamental interaction here  is the neutralino-quark
scattering. However, since the quarks are bound inside the nucleon and
the nucleons are bound inside the nucleus, one has to deal with 
nuclear physics effects.We shall return to these later. We consider first the
fundamental interaction involving neutralino and quarks. This interaction 
involves exchange of Z boson, squarks and Higgs. The dark matter 
in the Milky Way halo is estimated  to have a 
velocity of $v\sim 300 km/s$. In scattering involving these neutralinos
the typical kinetic energy exchange is O(1keV $M_{\chi_1}/1GeV)$. Because 
the energy exchanges are small one can approximate the pole exchange
diagrams by contact interactions. The total interaction is of the form
\begin{equation}
L_{eff}= L_{eff}^{(SD)}+ L_{eff}^{(SI)}
\end{equation} 
\noindent
where $L_{eff}^{(SD)}$  is a spin dependent interaction 
and   $L_{eff}^{(SI)}$ is a spin independent part. $L_{eff}^{(SI)}$
is given by 
\begin{equation}
L_{eff}^{(SD)}=({\bar{\chi}_1}{\gamma^{\mu}}\gamma^5\chi_1)[\bar{q}\gamma_{\mu}
(A_LP_L+A_RP_R)q]
\end{equation} 
\noindent 
where $P_{R,L}$ are projection operators with $P_{R,L}=(1\pm\gamma_5)/2$ etc.
Note that because of the Majorana nature of the neutralino 
$\bar\chi_1{\gamma^{\mu}}\chi_1$=0 and thus only the axial-vector
bilinear involving the neutralino enters. The spin-dependent interaction 
receives contributions from the Z exchange and from the chirality diagonal
(LL and RR) squark exchange terms. For the Z-exchange one finds that
$A_L^Z,A_R^Z \propto (n_3^2-n_4^2)$ which means that this exchange not only
depends on the Higgsino components of the neutralino [defined in Eq.(9)]
but on their 
mismatch as well. On the other hand the chirality diagonal squark exchange
depends on the  gaugino components of the neutralino, i.e., one has 

 \begin{equation}
A_L^{\tilde q}\propto (n_1g_2 T_3+n_2 g_1\frac{Y_L}{2})^2,~~A_R^{\tilde q}\propto
(n_2g_1\frac{Y_R}{2})^2
\end{equation} 
\noindent 
We note in passing that the spin dependent interaction discussed 
above  involves no interference 
between the gaugino components and the Higgsino components. 
This aspect of the spin dependent interaction is in contrast to the 
scalar interaction which we discuss next. The 
effective  scalar  interaction in given by  

\begin{equation}
L_{eff}^{(SI)}= (\bar{\chi}_1 \chi_1)(\bar{q}C m_qq) 
\end{equation} 
\noindent
 The scalar interaction receives contributions from Higgs exchange and
 from the squark exchange. 
  The contribution to C from the Higgs exchange obeys

\begin{equation}
C_{Higgs}\propto   {g_2^2\over 4M_W} 
 \left [{F_h\over m_h^2}
  \left\{  {cos\alpha_H\over sin\beta} \atop{-sin\alpha_H\over cos\beta}
           \right\}
            + 
  {F_H\over m_H^2} \left\{  {sin\alpha_H\over sin\beta} 
  \atop{cos\alpha_H\over cos\beta} \right\} \right ]{u-quark\atop d-quark}  
\end{equation} 
\noindent
where h is the light CP even Higgs, H is the heavy CP even Higgs,
$F_h =(n_1-n_2 tan\theta_W)$$(n_4 cos\alpha_H$ +$n_3 sin\alpha_H)$,
$F_H = (n_1-n_2tan\theta_W)$ $(n_4 sin\alpha_H-n_3 cos\alpha_H)$ and
$\alpha_H$ is the rotation angle needed to diagonalize the h-H$^{\circ}$ mass
matrix and is defined such that\cite{gunion}

\begin{equation}
cos(2\alpha_H)=\frac{M^H_{11}-M^H_{22}}{\sqrt{(M^H_{11}-M^H_{22})^2+
4M^{H2}_{12}}},~~tan2\alpha_H=\frac{2M^H_{12}}{(M^H_{11}-M^H_{22})}
\end{equation}
where $M_{ij}^H$ is the Higgs mass matrix in the CP even sector.
We emphasize here the fact that the inclusion
of loop corrections to the Higgs masses is important in the computation
of the Higgs mixing angle\cite{haber,an2}. Further, the contribution of the
heavy Higgs exchange is in general not small\cite{heavyh}. 
Thus while the heavy Higgs
contribution is indeed small in the  up quark sector, its contribution 
in the down quark sector can be very significant. The reason for this
is that in the down quark sector the light Higgs contribution is
proportional to $sin\alpha_H$ while the heavy Higgs contribution is 
proportional to $cos\alpha_H$. Now  over much of the parameter space 
one finds that $tan\alpha_H = O(1/10)$. This leads to  a suppression
in the light Higgs sector and  an enhancement in the heavy Higgs 
sector. Further, $n_3/n_4\sim tan\beta$ (see Appendix A) and hence $F_H$
can become significantly larger than $F_h$. These enhancements
 in the heavy sector
can sometimes even overcome the natural mass suppression of $m_h^2/m_H^2$ making the
heavy Higgs sector contribution quite large. Finally, the 
(chirality non-diagonal) squark exchange also contributes to the
scalar interaction and one has:

\begin{equation}
C_{\tilde q}\propto (\frac{n_1g_2T_3+n_2g_1\frac{Y_L}{2}}{(m_{\tilde q_L}^2
-m_{\chi_1}^2)}-\frac{n_2g_1\frac{Y_R}{2}}{(m_{\tilde q_L}^2
-m_{\chi_1}^2)})(n_3\frac {T_3+\frac{1}{2}}{sin\beta}-n_4\frac{T_3-\frac{1}{2}}
{cos\beta})
\end{equation} 
\noindent

We note  that the scalar interaction involves an interference of 
 the gaugino and the Higgsino components of the neutralino. Thus
 if  the lightest neutralino were to be strictly a Bino, the
 scalar interaction would vanish and the neutralino- nucleus scattering
 would be strictly governed by the spin dependent interaction. However,
 such a situation is never realised and the scalar 
 interaction always plays an important role in the event rate
 analysis. The scalar interaction is further enhanced by the fact that
  when one sums over all the quarks in 
 the nucleus, one 
produces in the case of the scalar interaction a  nuclear mass factor M$_N$,
which enhances the contribution of the scalar term.  
Thus the scalar interaction is the dominant one for heavy nuclei. 
The event rate for
a detector invoving a target nucleus of mass  $M_N$ is then given by

\begin{equation}
R=\left[ R_{SI}+R_{SD}\right ] \left [{\rho_{\tilde{\chi}_{1}}\over 0.3GeV
cm^{-3}}\right ] \left [{v_{\tilde{\chi}_{1}}\over 320 km/s}\right ]{events\over
kg~ da}  
\end{equation}
\noindent 
Here ${\rho_{\tilde{\chi}_{1}}}$ is the local mass density of
${\tilde{\chi}_{1}}$, and  $v_{\tilde{\chi}_{1}}$ is 
the incident ${\tilde{\chi}_{1}}$ 
velocity. The spin dependent (SD) rate is given by
\begin{equation}
R_{SD} = {16 m_{\tilde{\chi}_{1}}M_N\over \left [M_N+m_{\tilde{Z}_{1}}\right ]^2}
\lambda^2J(J+1)\mid A_{SD}\mid^2
\end{equation}
\noindent
where J is the nuclear spin and $\lambda$ is defined by $<N\mid\sum
{\stackrel{\rightarrow}{S}_i}\mid N>=\lambda<N\mid{\stackrel{\rightarrow}
{J}}\mid N>$.  For the spin independent part (SI) one has 
\begin{equation}
R_{SI}={16m_{\tilde{\chi}_{1}}M_N^3M_Z^4\over\left
[{M_N+m_{\tilde{\chi}_{1}}}\right ]^2}{\mid A_{SI}\mid^2}  
\end{equation}
 We note that for large M$_N$, $R_{SD}\sim 1/M_N$, while 
 R$_{SI}\sim M_N$ which implies that heavier the  nucleus the more 
 sensitive it is to the spin independent scattering.\\
 
    We discuss  now the uncertainties associated with the computation
    of $R_{SI}$ and $R_{SD}$. In the computation of $R_{SI}$ most
    of the uncertainty is associated with the computation of the 
    matrix elements of the operator $m_q \bar q q$  between the nucleon
    states. One may write the matrix elements in the form 
    \begin{equation}
     <n|m_q \bar q q|n>=m_n f_q^{(n)} 
    \end{equation}

\noindent	
    Estimates give $f_u \approx f_d\approx 0.05 $, $f_s\approx 0.2$ 
    with as such much as $50\%$ uncertainty in their 
    determination\cite{cheng}. 
    	Further  the heavy quark contributions cannot be ignored. 	
    		They couple to nucleons at the loop level via exchange
    		of gluons and contribute to the mass of the nucleon.
                One may write     $<n|m_Q$ $\bar QQ|n>$=$\frac{2}{27}$$m_n$
                (1- $\sum _q $$f_q^{(n)})$. An uncertainty also exists in
                the determination of $R_{SD}$. This quantity depends on
                the matrix element
	
\begin{figure*}
\begin{center}
\hspace*{0.3in}
\psfig{figure=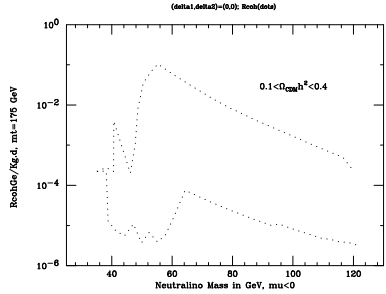,height=1.5in}
\end{center}
\caption[]{ Maximum and minimum of event rates for Ge with scalar interaction.}
\label{fig:1}
\end{figure*}

\begin{equation}
<n\mid{\bar{q}}\gamma^{\mu}\gamma^5q\mid n>=2s^{\mu}_{(n)}\Delta q
\end{equation}
Here $s_{(n)}^{\mu}$ is the spin 4-vector of the nucleon, and $\Delta q$ 
is the quark polarizability. Currently there are two determinations of 
 $\Delta q$ , one using the older EMC data\cite{emc}, and the other using the
 new SMC data\cite{ellis2}. The analyses using  the EMC
   data and the SMC data
 give

\begin{equation}
\Delta u = 0.7\pm0.08, \Delta d=-0.49\pm0.08, \Delta s=-0.15\pm 0.08 (EMC)
\end{equation}

\begin{equation}
\Delta u =0.83\pm 0.03,  \Delta d =-0.43\pm 0.03,  \Delta s=-1.0\pm 0.03 (SMC)
\end{equation}
Now the variations in $\Delta q$, and specifically the variation in 
$\Delta s$ can give rise to a significant variation in $R_{SI}$. 
However,  $R_{SI}$ is typically not a large  component of the total R
except  for light nuclei such as CaF$_2$. Thus  the event rate analyses 
in lighter nuclei are the ones most sensitive to variations in the
values of $\Delta q$ while the event rates in heavier nuclei are not
very sensitive.     
                  
\section{Analysis of Event Rates for Universal Soft SUSY Breaking}
We discuss now the analysis of the event rates. Here we give the
analysis under the assumption that the soft SUSY breaking parameters
are universal at the GUT scale. As discussed in Sec 2, 
the parameter  space of supergravity grand unification in this case 
 is 4 dimensional and is spanned by  
\begin{equation}
m_0, m_{\tilde g}, A_t, tan\beta, sign(\mu)
\end{equation}
As mentioned already in sec 2 the condition that $\tilde t_1$ not
turn tachyonic places severe constraints on the parameter space. From 
eqs(12) and (13) one may see this as follows:
for $A_t$ negative  $A_R^2$ becomes large quickly since the $A_t$ term
and the gluino term reinforce each other driving the $\tilde t_1$ 
tachyonic. For $A_t$ positive one finds
that $A_t$ also cannot get too large since that would make $A_R^2$ large 
($m_{\tilde g}<1 TeV$)
and again drive the light stop tachyonic. Numerical analysis shows that the 
no tachyonic condition limits $A_t$ to the following range\cite{nwa,an3}

\begin{equation}
-1.0~{\stackrel{<}{\sim}}~ A_t/m_o~{\stackrel{<}{\sim}}~ 5.5
\end{equation}
\noindent
Thus a heavy top mass  eliminates a large amount of the SUSY
parameter space.

\begin{figure*}
\begin{center}
\hspace*{0.3in}
\psfig{figure=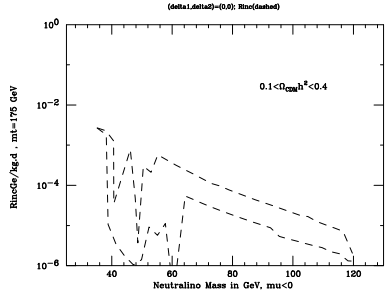,height=1.5in}
\end{center}
\caption[]{ Maximum and minimum of event rates for Ge with spin dependent
 interaction.}
\label{fig:2}
\end{figure*}
Event rates are computed under the constraint of the naturalness
assumption on $m_0, m_{\tilde g}\leq$ 1TeV, and $\tan\beta < 25$.
$A_t$ is then restricted by the electro-weak symmetry breaking and
by the  condition that there be no tachyons,i.e., eq(38). In this range of the 
parameter space we compute the event rates for various target materials.
The maximum and the minimum of the event rates for  Ge are exhibited in 
Figs 1-3. Fig.1 is for coherent scattering when we take
into account only the scalar part of the interaction. Fig.2 gives the 
analysis of incoherent part of the event rates where we take into 
account only the spin dependent part. Fig.3 gives a composite of the
event rates including the coherent part, the incoherent part and the
total. 
A comparison of the maximum and the minimum curves shows that for 
the case of Ge(a heavy nucleus) the maximum event rates 
are almost 100\% dominated
by the coherent part, while the minimum curves have a substantial 
component( as much as 40\%) of incoherent part above a neutralino
mass of about 60 GeV. Fig.4 shows the corresponding curves for a light 
nucleus, CaF$_2$. We also note that both the maximum and  minimum  
 event rates
 show a sharp dip
in the vicinity of the Z-pole. This phenomenon arises as a consequence
of the accurate method of computing the relic density which takes 
account of the correct thermal averaging over the Z and the Higgs poles
in the annihilation of the neutralinos(the lack of a second sharp dip in the
event rates from the Higgs pole is due to the fact that one is 
integrating over the parameter space which smears  the Higgs mass pole effects).
The dependence of the event rates on the neutralino mass beyond the
annihilation region involving the Z and the Higgs poles can also be 
understood. The event rates in this region are controlled by $\mu$.
 Since $\mu^2$ is an
increasing function of -m$_{H_{2}}^2$, by Eq. (7) it is an increasing
function of m$_{\tilde{\chi_1}}$ and m$_o$. As discussed above the 
maximum event rates are dominted by A$_{SI}$  which 
is proportional to ($n_2n_3$) [or ($n_2n_4)$] which by Eqs. (48-51)
are the leading [O(M$_Z/\mu$] terms.  These decrease as $\mu$ increases. 
Thus we expect R to decrease as m$_{\tilde{\chi_1}}$ and m$_o$ increase. 
The analysis of Fig 3 bears this out. Finally we mention that the 
the maximum curve in the event rates has a strong dependence on tan$\beta$.
It increases as tan$\beta$ increases. This can be seen from Eq.(27) 
where the leading d-quark contribution has $C_{Higgs}\sim $$
1/cos\beta$$\sim$$tan\beta$.

 \section{ Effects of Non-universal Soft Breaking}

 Most of the previous analyses in supergravity grand unification have been
done using  universal boundary conditions for the soft SUSY breaking masses
at the scale $M_G$. However, the framework of supergravity unification 
allows for the presence of non-universalities from a general Kahler 
potential~\cite{soni}.  In the present analysis
we wish to examine the effect of non-universalities on the 
predictions on  dark matter event rates. In the present analysis for 
specificity we shall confine
ourselves to consideration of non-universalities only in the 
Higgs sector\cite{matallio,olech,polon,berez}.
Thus we assume that the only non-universalities we have at the GUT scale
 $M_G$ are given by
~\\
\begin{equation}
m_{H_1}^2 =  m{_0^2} (1 + \delta_1) ; ~~m_{H_2}^2 = m{_0^2} (1 + \delta_2),
\end{equation}
\noindent
where  $\delta_{1,2}$ represent
the deviation from the universal reference mass $m{_0}$. All
the remaining parameters are assumed universal. 
The important parameters that enter in the low energy domain are 
  $\mu^2$, m$_A^2$,
and m$_{\tilde t_{1,2}}$. The role of m$_{\tilde t_{1,2}}$
has already been discussed in Sec.2 We exhibit here the effects of
non-universalities on the parameters 
$\mu^2$ and  $m_A^2$. We find 

\begin{figure*}
\begin{center}
\hspace*{0.3in}
\psfig{figure=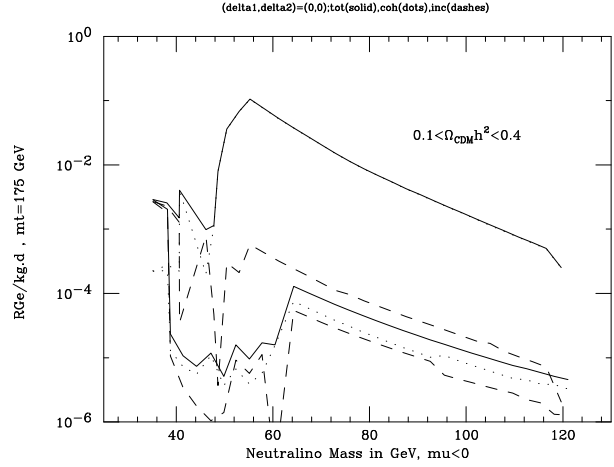,height=2.4in}
\end{center}
\caption[]{A comparison of the maximum and the minimum of event rates for Ge 
for three cases:(i) the coherent part arising from the scalar interaction
(dotted),(ii) the incoherent part arising from the spin dependent 
interaction(dashed), and (iii) sum of cases(i) and (ii) (solid). The maximum
dotted line overlaps the maximum solid line( see Fig 1 for comparison
of the dotted and solid lines)}

\label{fig:3}
\end{figure*}
\begin{equation}
\mu^2=m_0^2 C_1+A_0^2 C_2 +m_{\frac{1}{2}}^2C_3+A_0m_{\frac{1}{2}}C_4
-\frac{1}{2}M_Z^2
\end{equation}
Here 
\begin{equation}
C_1=\frac{1}{t^2-1}(1-\frac{3 D_0-1}{2}t^2)+
\frac{1}{t^2-1}(\delta_1-\frac{ D_0+1}{2}\delta_2 t^2)
\end{equation}

\begin{equation}
C_2=-\frac{t^2}{t^2-1}k, 
~C_3=-\frac{1}{t^2-1}(g- t^2 e), 
~C_4=-\frac{t^2}{t^2-1}f 
\end{equation}
 where $t\equiv tan\beta$ and where the functions e,f,g,k are as defined 
in ref(10). Similarly, for $m_A^2$ one has

\begin{figure*}
\begin{center}
\hspace*{0.3in}
\psfig{figure=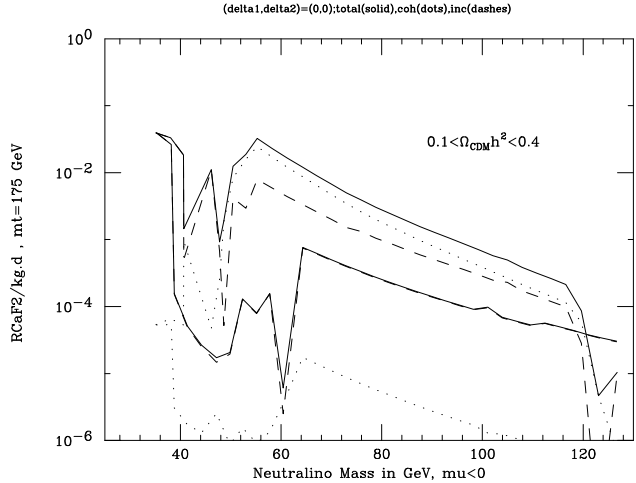,height=2.5in}
\end{center}
\caption[]{ A comparison of the maximum and minimum of event rates for $CaF_2$ 
for three cases:(i) the coherent part arising from the scalar interaction
(dotted),(ii) the incoherent part arising from the spin dependent 
interaction(dashed), and (iii) sum of cases(i) and (ii) (solid)}

\label{fig:4}
\end{figure*}

\begin{equation}
m_A^2=m_0^2 D_1+A_0^2 D_2 +m_{\frac{1}{2}}^2D_3+A_0m_{\frac{1}{2}}D_4
-\frac{1}{2}M_Z^2
\end{equation}
where 
\begin{equation}
D_1=\frac{t^2+1}{t^2-1}(1-\frac{3 D_0-1}{2})+
\frac{t^2+1}{t^2-1}(\delta_1-\frac{ D_0+1}{2}\delta_2 t^2)
\end{equation}

\begin{equation}
D_2=-\frac{t^2+1}{t^2-1}k, 
~D_3=-\frac{t^2+1}{t^2-1}(g- e), 
~D_4=-\frac{t^2+1}{t^2-1}f 
\end{equation}
In the above $D_0$ defines the Landau pole in the top quark Yukawa coupling $y_0$, 
i.e,
\begin{equation}
y_0=\frac{y_t}{E(t)D_0}; ~D_0=1-6 y_t \frac{F(t)}{E(t)}
\end{equation}
and 
\begin{equation}
E(t)=(1+\beta_3)^{\frac{16}{3b_3}}(1+\beta_2)^{\frac{3}{b_2}}
(1+\beta_1)^{\frac{13}{9b_1}}
\end{equation} 
where $\beta_i=\alpha_i(0)b_i/4\pi$, $\alpha_i(0)$ are the gauge
coupling constants at the GUT scale with 
$\alpha_1=(5/3)\alpha_Y$, 
$b_i$ are 
the one loop
beta function co-efficients defined  by $(b_1,b_2,b_3)=
(33/5,1,-3)$, 
and $F(t)=\int_0^t E(t)dt$. 
 
Normally we expect $|\delta_{1,2}|\leq 0.5$, and a similar 
level of non-universality in the relevant mass spectra at 
the electro-weak scale.However, under special
circumstances enhancement of non-universalities can occur when the 
universal terms are suppressed.
 To exhibit this 
phenonmenon we consider the solution for $\mu^2$. The dominant terms 
here are the first two terms, i.e., the terms which involve $C_1$
and $C_2$. 
Now $C_1$ contains a universal part and a non-universal part. 
For large tan$\beta$, the universal part accidentally cancels when 
$D_0$=1/3 which occurs for a value of $m_t\approx 168$ GeV, 
and consequently the non-universality effects get enhanced.
An enhancement of non-universalities also occurs from the $A_0C_2$ 
term, but in a somewhat different manner. Here the non-universalities
are enhanced when  the residue of the Landau pole in $A_0$ vanishes.
This happens when 
$A_t\approx$0.6 $m_{\tilde g}$.

\begin{figure*}
\begin{center}
\hspace*{0.3in}
\psfig{figure=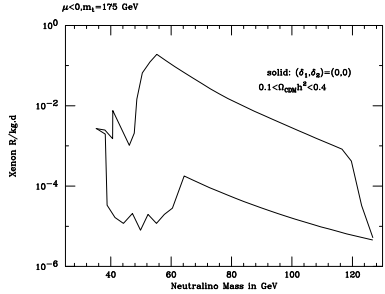,height=1.5in}
\end{center}
\caption[]{ Maximum and minimum of event rates for Xenon for the universal
case when $\delta_1=0=\delta_2$}
\label{fig:5}
\end{figure*}
 \section{ Event Rates With Non-universal Soft SUSY Breaking}

In this section we analyse the event rates for Xenon with
both universal as well as with non-universal soft SUSY breaking terms.
Fig.5 exhibits the maximum and the minimum of event rates for Xenon
for the case  $\delta_1=0=\delta_2$ . The general features of this
case are  similar to the case of Ge discussed earlier and the observations
made there apply here as well. Next we discuss the non-universal case. 
When $\delta_2$ $>$ 0 and $\delta_1$ $<$ 0, the non-universal terms in 
Eq.(40)
decrease $\mu^2$  and can in some cases drive  $\mu^2$ 
negative, eliminating such points from the parameter space.
This effect is larger for small tan$\beta$.  Fig.6 exhibits this case for
$\delta_2$ = -$\delta_1$ = 1 where the major effect is to significantly raise
the minimum event rate (for lower values of m$_{\tilde \chi{_1}}$) since it is 
the smallest tan$\beta$ that gives rise to the minimum values of R.
Fig.7 exhibits the opposite effect, i.e., a decrease in minimum value of R,
when $\delta_2=-1$, $\delta_1=1$ . The decrease in the minimum arises 
since $\mu^2$ is increased and R is a falling
function of $\mu$. Fig 8 gives for comparison 
the composite of the three cases discussed above. We note that as
 $m_{\chi_1}$ increases and hence $m_{\tilde g}$ increases  one finds that
 $A_R$ increases and hence the Landau pole contribution becomes more
dominant. In this region the effect of the non-universalities tends
to get washed out. Further the effect of the Z-pole disappears for 
$m_{\chi_1}\geq 65$ GeV. This explains in part why the minimum and the maximum 
event rates in the region of large gluino masses are insensitive to the
effect of  the non-universalities. 
Also, as  m$_t$ is increased, D$_0$ becomes smaller and the Landau pole term in Eq.
(40) dominates, making the effects of non-universal terms again 
less significant.  
Numerical analysis supports this conclusion.

\begin{figure*}
\begin{center}
\hspace*{0.3in}
\psfig{figure=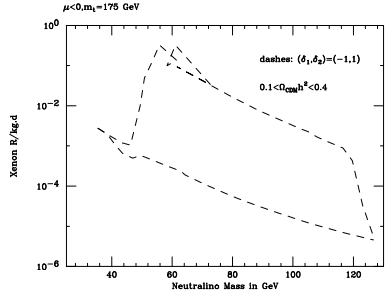,height=1.5in}
\end{center}
\caption[] { Maximum and minimum of event rates for Xenon for the non-universal
case when $\delta_1=-1=-\delta_2$}
\label{fig:6}
\end{figure*}

 Finally we discuss the consequences of a more accurate
 determination of $\Omega h^2$. One expects such a situation to occur
 in the next round of COBE like sattelite experiments which may determine
 the Hubble parameter to  within 10\%  or better accuracy.This determination of
 the Hubble parameter will strongly reduce the error in $\Omega h^2$. Here
 we analyse what such an accurate determination will do to the prediction 
 of the neutralino event rates. For specificity let us assume that 
 $\Omega h^2$ is constrained to lie in the narrow band
   $0.225<\Omega h^2<0.275$. The narrowing of the range of $\Omega h^2$
   more stringently limits the parameter space of the SUGRA model.
   The result of the event rate analysis  for Xenon under the above 
   constraint is shown in Fig.9.
   An interesting new phenomenon is the appearance of  peaks in the
   minimum curves. These sharp peaks originate from the constraints to
   prevent charge and color breaking minima\cite{ccb}(CCB constraints). 
   The position of 
   the peak is 
   determined by the mass parameter $m_{H_2}^2$. The peak moves to the 
   right as   $m_{H_2}^2$  increases and moves to the left as 
    $m_{H_2}^2$ decreases. The first situation occurs for the case 
    $\delta_1<0, \delta_2>0$ , and the second for the case 
     $\delta_1>0, \delta_2<0$. Numerical analysis for these cases as
     given by the dashed curve and the dotted curve support the analytical
     result.  Comparison of Figs.8 and 9 shows that a 
     narrower range in $\Omega h^2$ more sharply  constrains the SUGRA parameter space
   	and the allowed band of event rates shrinks.

\begin{figure*}
\begin{center}
\hspace*{0.3in}
\psfig{figure=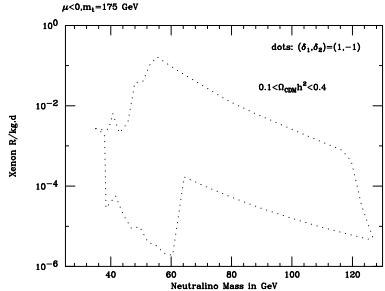,height=1.5in}
\end{center}
\caption[]{ Maximum and minimum of event rates for Xenon for the non-universal
case when $\delta_1=1=-\delta_2$}
\label{fig:7}
\end{figure*}

\section{CONCLUDING REMARKS}

In this review we have given an analysis of event rates for the observation
of dark matter in terrestrial dark matter detectors. 
 Supergravity GUT models under the assumption of R parity invariance 
 predict the neutralino to be cold dark matter over most of the parameter
 space of the model. The analysis of the event rates is carried out
 by ranging over the full parameter space of the SUGRA model 
 within the prescribed naturalness constraints. Then we finds that
 the event rates can range from 
  R = (10$^{-5}$ to $\approx$ 5) events/kg da. We find that for  a heavy
  target such as Ge or Xenon the maximum event rates are almost totally 
  coherent(which is controlled by the scalar interaction) while
  the minimum event rates can have a significant incoherent component
  (which is controlled by the spin dependent interaction). For a light 
  target such as CaF$_2$, one finds that the minimum event rates are 
  essentially incoherent while the maximum event rates have  a very
  significant component of coherent scattering. 
  
 We  have also analysed in this paper the effects of non-universal soft 
SUSY breaking parameters in the Higgs sector on event rates. 
It is found that the 
non-universal soft breaking effects  are most significant
in the neutralino mass range m$_{\tilde\chi{_1}}~\simeq$ (35--65) GeV.
In this region one finds that the minimum event rates can be decreased or
increased by a factor of 10 depending on the nature of non-universality.

  The above analyses show that experiments with a sensitivity of 
  $R\geq$ 0.005 events/kg da can explore a significant part of the
 parameter space.  However, a further significant improvement
in sensitivity by a factor of about 100 or more would be needed to explore the
entire parameter space. For the case of non-universalities with 
 $\delta_2=-\delta_1=1$  there is an improvement by  a factor of
 about 10  in the event rates in the energy domain $M_{\tilde Z_1}\leq 65$ GeV.
 However, for the case $\delta_2=-\delta_1=-1$ one finds a reduction in the 
 event rates by a factor of about 10 in the same mass region.
  Thus whether or not the event rates are enhanced  or reduced depends 
  on the nature of the non-universality.

\begin{figure*}
\begin{center}
\hspace*{0.3in}
\psfig{figure=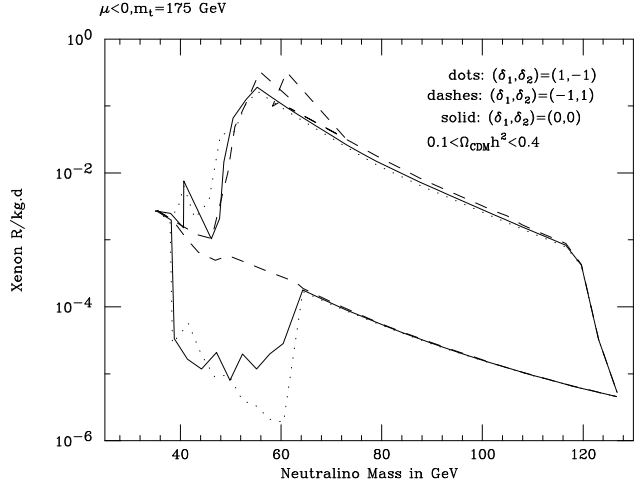,height=2.5in}
\end{center}
\caption[] {Comparison of the maximum and minimum of event rates for Xenon 
for three cases:(i) universal case when $\delta_1=0=\delta_2$(solid);
(ii)(i)non-universal case when $\delta_1=-1=-\delta_2$(dashed);
(iii) non-universal case when $\delta_1=1=-\delta_2$(dotted)}
\label{fig:8}
\end{figure*}

The above results have important implications for dark matter detectors.
Thus the analysis shows that the parameter space of the model  
with $\delta_2=-\delta_1=1$ can be exhausted completely in the mass 
domain  $M_{\tilde Z_1}\leq 65$ GeV by an improvement in the detector
sensitivity by a factor of about 100. However, much greater sensitivity
will be needed to do the same for the case $\delta_2=-\delta_1=1$.
 Thus new detection techniques which have the potential of dramatically
 improving the detection sensitivities \cite{collar,cline}, 
 are important for  a complete exploration of
the parameter space of  the  supergravity unified model.

\begin{figure*}
\begin{center}
\hspace*{0.3in}
\psfig{figure=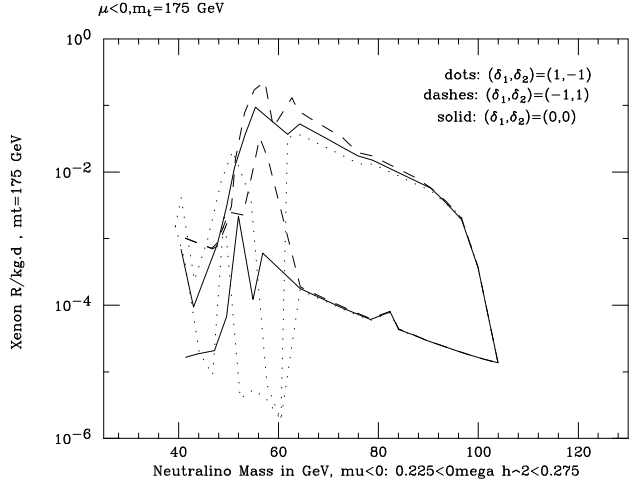,height=2.5in}
\end{center}
\caption[] {Comparison of the maximum and minimum of event rates for Xenon 
for three cases when $0.225<\Omega h^2<0.275$  
:(i) universal case when $\delta_1=0=\delta_2$(solid);
(ii)non-universal case when $\delta_1=-1=-\delta_2$(dashed);
(iii) non-universal case when $\delta_1=1=-\delta_2$(dotted)}
\label{fig:9}
\end{figure*}

\section*{Acknowledgments}
This research was supported in part by NSF grant numbers 
PHY-96020274 and PHY-9411543.

\section*{Appendix A}
 To second order perturbation theory one finds~\cite{an3} when
 $(M_Z/\mu)^2<<1$:

\begin{eqnarray}
n_{1} & \cong &
-{1\over2}{M_{Z}\over\mu}{1\over{(1-\tilde{m}_{1}^{2}/\mu^{2})}}{M_{Z}\over
{\tilde{m}_{2}}-\tilde{m}_{1}}
sin2\theta_{W}\left[sin2\beta +{\tilde{m}_{1}\over\mu}\right]\\
n_{2} & = &1-{1\over2}{M_{Z}^{2}\over\mu^{2}}{1\over{(1-\tilde{m}_{1}^{2
}/\mu^{2})^{2}}}
sin^{2}\theta_{W}\left[1+{\tilde{m}_{1}\over\mu} sin2\beta +{{\tilde{m}_{1}^{2}}
\over
\mu^{2}}\right]\\
n_{3} & = &{M_{Z}\over\mu}{1\over{1-\tilde{m}_{1}^{2}/\mu^{2}}}
sin\theta_{W}sin\beta\left[1+{\tilde{m}_{1}\over\mu} ctn\beta\right]\\
n_{4} & = &-{M_{Z}\over\mu}{1\over{1-\tilde{m}_{1}^{2}/\mu^{2}}}
sin\theta_{W} cos\beta\left[1+{\tilde{m}_{1}\over\mu} tan\beta\right]
\end{eqnarray}
\noindent
Eqs. (48-51) differ from the numerical computer results by amounts $\delta
n_{i}\stackrel{<}{\sim} 0.03$ over a large part of the parameter space and
 are often significantly better.

\section*{Appendix B}

As discussed in Sec 3, the conventional approximation in the computation of
relic density has been the expansion of $\sigma v$ in a power series 
expansion in v. However, in the annilation of the neutralinos one
encounters Z and Higgs poles. A power series expansion of $\sigma v$ in
this case breaks down. One must therefore treat the pole correctly in
the thermal averaging and in the computation of $J(x_f)$\cite{na0}.
We illustrate this below for the case of the Higgs pole.  Here one has

\begin{eqnarray}
\sigma v=\frac{A_{Higgs}}{m_{\tilde\chi_1}^4}\frac{v^2}{((v^2-\epsilon_h)^2
+\gamma_h^2))}\\
\epsilon_h=(m_h^2-4 m_{\chi_1}^2)/m_{\chi_1}^2 \\ 
\gamma_h=m_h\Gamma_h/m_{\chi_1}^2  
\end{eqnarray}
In the above $m_h$ is the Higgs mass and $\Gamma_h$ is the Higgs width.
 $J(x_f)$ involves a double integration over a 
pole which for accuracy requires a high precision numerical computation.
It is possible to significantly simplify the expression for  $J(x_f)$
by carrying out one of the integrations analytically. One may then 
write $J(x_f)$ in the following  form

\begin{equation}
J_{Higgs}(x_f)=\frac{A_{Higgs}}{2\sqrt 2 m_{\tilde\chi_1}^4}
(I_{1h}+\frac{\epsilon_h}{\gamma_h} I_{2h})
\end{equation}

\begin{equation}
I_{1h}=\frac{1}{2}\int_0^{\infty}dyy^{-\frac{1}{2}}e^{-y}
log[\frac{(4yx_f-\epsilon_h)^2+\gamma_h^2}{\epsilon_h^2+\gamma_h^2}]
\end{equation}

\begin{equation}
I_{2h}=\frac{1}{2}\int_0^{\infty}dyy^{-\frac{1}{2}}e^{-y}
[tan^{-1}(\frac{(4yx_f-\epsilon_h)^2+\gamma_h^2}{\gamma_h})
+tan^{-1}(\frac{\epsilon_h}{\gamma_h})]
\end{equation}
The numerical analysis of $J(x_f)$ is significantly simplified using the
 reduced  form. A  similar treatment of the Z pole term can also be
 carried  out.

\end{document}